# Simultaneous Superoscillations in Space and Time in Nonseparable Light Pulses

Yijie Shen[1,2], Nikitas Papasimakis[3], Nikolay I. Zheludev[3,4]

1. Centre for Disruptive Photonic Technologies, School of Physical and Mathematical Sciences, Nanyang Technological University, Singapore 637371, Singapore
2. School of Electrical and Electronic Engineering, Nanyang Technological University, Singapore 639798, Singapore
3. Optoelectronics Research Centre & Centre for Photonic Metamaterials, University of Southampton, Southampton SO17 1BJ, UK
4. Institute for Advanced Study, Texas A&M University, USA

**Abstract**
A remarkable phenomenon of superoscillations implies that electromagnetic waves can locally oscillate in space or time faster than the fastest spatial and temporal Fourier component of the entire function. This phenomenon allows to focus light into an arbitrary small hotspot enabling superresolution imaging and optical metrology with accuracy far beyond the Abbey-Reileigh diffraction limit. Here we show that, in band-limited supertoroidal light pulses, the temporal and spatial superoscillations can be observed simultaneously at a specific region in space and at a specific interval in time.

**Introduction**
In the last few years optical superoscillations attracted a considerable attention due to their fascinating electromagnetic properties and applications in imaging and metrology [1,2]. With coherent monochromatic light, superoscillations can be generated in one, two or three dimensions [3,4], while the concept of superoscillations to an arbitrary number, $D$, of spatial dimensions was developed by Berry and Dennis [5], whereby a monochromatic wave satisfying the Helmholtz equation $\nabla^2\psi(\mathbf{r}) + k_0^2\psi(\mathbf{r}) = 0$, with $\mathbf{r}=\{x_1, …, x_D\}$ and $k_0=\omega/c$ the free-space wavenumber, contains local regions that oscillate faster than $k_0=\omega/c$. Recently, superoscillations have been considered in the temporal domain resulting in counterintuitive light-matter interactions and high resolution spectroscopy [6,7].

Here we demonstrate that spatial and temporal superoscillations can be observed simultaneously at a given point in time and space, in the recently introduced family of spatiotemporally structured supertoroidal pulses (STPs) [8]. In contrast to conventional pulses, STPs cannot be expressed as products of spatial and temporal functions, but instead they exhibit nonseparability in the space and time domains resulting in a complex topological spatiotemporal structure [8]. As a result of their nonseparable nature, STPs cannot be presented in the form satisfying the Helmholtz equation. A characteristic member of this family of pulses is the toroidal light pulse or "Flying Doughnut", the focused finite-energy single-cycle solution of toroidal topology predicted by Hellwarth and Nouchi in 1996 [9] and recently observed experimentally from optical to microwave domains [10-12].

**Results**
Generally, an STP can be characterized by two length parameters, $q_1$ and $q_2$, representing the dominant wavelength and Rayleigh range, respectively, and a real dimensionless parameter $\alpha$, which controls the topology of the pulses, while $\alpha\geq1$ ensures finite pulse energy [8]. STPs can be either azimuthally (TE) or radially polarized (TM). With no loss of generality, here we focus on the TE case. Such a pulse can be represented by its instantaneous azimuthal electric field

$E_\theta(r,z,t)=A(r,z,t)e^{i\phi(r,z,t)}$, where $A$ and $\varphi$ correspond to the amplitude and phase. The azimuthal electric field for a free-space propagating STP is given by (also see supplementary, S1):

$$E_\theta^{(\alpha)} = -2\alpha i f_0 \sqrt{\frac{\mu_0}{\varepsilon_0}} \left\{ \frac{(\alpha+1)r(q_1+i\tau)^{\alpha-1}(q_1+q_2-2ict)}{\left[r^2+(q_1+i\tau)(q_2-i\sigma)\right]^{\alpha+2}} - \frac{(\alpha-1)r(q_1+i\tau)^{\alpha-2}}{\left[r^2+(q_1+i\tau)(q_2-i\sigma)\right]^{\alpha+1}} \right\} \quad (1)$$

$$H_r^{(\alpha)} = 2\alpha i f_0 \left\{ \frac{(\alpha+1)r(q_1+i\tau)^{\alpha-1}(q_2-q_1-2iz)}{\left[r^2+(q_1+i\tau)(q_2-i\sigma)\right]^{\alpha+2}} - \frac{(\alpha-1)r(q_1+i\tau)^{\alpha-2}}{\left[r^2+(q_1+i\tau)(q_2-i\sigma)\right]^{\alpha+1}} \right\} \quad (2)$$

$$H_z^{(\alpha)} = -4\alpha f_0 \left\{ \frac{(q_1+i\tau)^{\alpha-1}\left[r^2-\alpha(q_1+i\tau)(q_2-i\sigma)\right]}{\left[r^2+(q_1+i\tau)(q_2-i\sigma)\right]^{\alpha+2}} + \frac{(\alpha-1)(q_1+i\tau)^{\alpha-2}(q_2-i\sigma)}{\left[r^2+(q_1+i\tau)(q_2-i\sigma)\right]^{\alpha+1}} \right\} \quad (3)$$

where $(r, \theta, z)$ are cylindrical coordinates, $t$ is time, $c = 1/\sqrt{\varepsilon_0\mu_0}$ is the speed of light, and $\varepsilon_0$ and $\mu_0$ are the permittivity and permeability of vacuum. where $f_0$ is a normalizing constant, $\tau = z - ct$, $\sigma = z + ct$, $q_1$ and $q_2$ are parameters with dimensions of length and act as effective wavelength and Rayleigh range under the paraxial limit, while $\alpha$ is a real dimensionless parameter as supertoroidal index that must satisfy $\alpha \geq 1$ to ensure finite energy solutions. Whereas STPs are not bandlimited pulses, here we consider their versions with time domain and space domain spectra truncated at $\omega<\omega_m$ and $k<k_m=\omega_m/c$. Truncated STPs retain all the main features of their non-bandlimited counterparts for $\omega_m \gg 4c/q_1$ (see details in Supplementary Material S2), including collocated spatial and temporal superoscillations.

We illustrate the presence of simultaneous space-time superoscillations by considering an STP with $\alpha=50$, $q_2=50q_1$, $q_1=1$ (see Fig. 1). The spatial variation of the electric field along the radial direction at $(z=0, t_s)$ is presented in Fig. 1c, where we observe that a segment in the off-axis region $(r\sim r_s)$ oscillates considerably faster than the harmonic oscillation of the maximal radial frequency $k_m$ (dashed red line). Similarly, in Fig. 1d, we show the temporal variation of the electric field at $(z=0, r_s)$, which around $t\sim t_s$ exhibits oscillations faster than the harmonic oscillation of maximal temporal frequency $\omega_m$ (dashed red line).

We illustrate the emergence of STSOs in STPs by considering their spatial and temporal distribution of the electric field at focus $(z=0)$ for a fixed cut-off frequency $\omega_m=ck_m=2c/q_1$. Figures 2(a1-a4) show the distribution of the logarithm of the electric field, i.e. $\log_{10}|\text{Re}[E_\theta(r,z=0,t)]|$, unwrapped phase of $\phi(r,z=0,t)$, large phase gradient, energy and Poynting vector distributions of the fundamental toroidal pulse $(\alpha=1)$. In Fig. 2(a3), there is only a small region with temporal local wavevector $\partial\phi/\partial r$ exceeding $k_m$, but no spatial SO and thus no STSO. This temporal SO is induced by the rapid cycle switching of the focused single-cycle pulse. In Fig. 2(a4), we also highlight the region of energy backflow accompanied by low amplitude values, in accordance with prior works [8,13]. Higher values of $\alpha$ lead to increasingly complex STPs with multi-cycle structure and dramatic spatiotemporal evolution, which results into an extreme spatiotemporal focusing effect with STSO. Figures 2(b1-b4) show the corresponding distribution of the logarithm of the electric field, unwrapped phase, large phase gradient regions, energy and Poynting vector distributions, for an STP with $\alpha=50$. Figure 2(b3) shows the presence of spatial SOs, where the local wavevector exceeds $k_m$, and temporal SOs, where the local temporal frequency exceeds $\omega_m$. Importantly, the spatial and temporal SOs overlap significantly indicating the presence of STSOs. We can also clearly observe that STSOs appear at low amplitude regions surrounded by higher-amplitude regions. Similarly, to the recently observed spatial SOs in the optical domain [13], STSOs are also accompanied by areas of energy backflow, see inset to Fig. 2(b4).

Figures 3(a) and 3(b) show temporal local wavevector profiles at specific radius $r=10q_1$ and spatial local wavevector profiles at specific time $t=2q_1/c$, respectively, for a series of STPs with varying value of parameter $\alpha$. We observe that the spatial and temporal rapid oscillations become increasingly stronger with increasing $\alpha$. Importantly, the observed oscillations carry a significant fraction of the total energy of the pulse. Figure 3(c) shows the ratio between energy within the rapid oscillations region, $E_{STSO}$, and the total energy of the pulse, $E_T$, as a function of parameter $\alpha$. We observe that for $\alpha>38$, STSO behaviour emerges with the energy of STSOs monotonically increasing with increasing $\alpha$, see Fig. 3(c).

Finally, we demonstrate that the spectra of STSO segments of STPs contain components out of the light cone, i.e. they oscillate faster than what would be expected based on the pulse bandwidth. Figure 4 shows the local spectra and entire spectra of STPs of different order, $\alpha = 1, 10, 50,$ and $100$, defined as the Fourier transform of the local space-time segment $E_\theta(r\sim r_s, z=0, t\sim t_s)$ into the $(k_r,\omega)$ domain. The spectra of full pulses are confined on the surface of the light cone, whereas local spectra present spectral components off the light cone. The presence of the off-light-cone components directly reveals the presence of STSOs. Whereas for the fundamental toroidal pulse ($\alpha=1$), local spectra are fully contained within the light cone, in the case of the STPs ($\alpha>1$) the off-cone components become stronger with increasing value of $\alpha$.

**Discussion**
Our work extends the concept of SOs to the space-time domain in the form of STSOs, extreme spatial and temporal SOs emerge at a specific location in space and at a specific moment in time and may diffuse as time passing by. We prove the existence of the STSO effect in space-time structured electromagnetic pulses, such as the STPs, by demonstrating the spatiotemporal analog of all characteristic features of SO waves.

From a practical standpoint, STSOs can be observed by generating STPs in the optical, THz and microwave ranges, following recent work on the generation of space-time non-separable toroidal pulses [10-12]. For instance, in the optical domain, commercially available Ti:Sa lasers can provide up to 200 nm bandwidth (FWHM) central at 800 nm corresponding to 6 fs long pulses. The corresponding limits to spatial and temporal focusing are ~400 nm and ~4.64 fs, respectively. Generating STPs with such a bandwidth would allow for STSO with spatial and temporal confinement of ~80 nm and 0.1 fs, exceeding the conventional spatial and temporal resolutions by 5 times and 45 times, respectively.

In summary, we have expanded the concept of superoscillations to include spatiotemporal effects and have demonstrated that the recently introduced STPs exhibit in principle arbitrarily small transient spatiotemporal focal spots. Our results may lead to applications in metrology, ultrafast spectroscopy, microscopy beyond the conventional spatial and temporal resolution limits. We argue that STSOs are not unique to the STPs considered here and we anticipate that STSO behavior may be found in other spatiotemporally structured light fields. Finally, while our discussion is limited to electromagnetic waves, different types of waves (e.g. acoustic, gravitational, or water waves) are expected to exhibit STSO phenomena.

# Figures

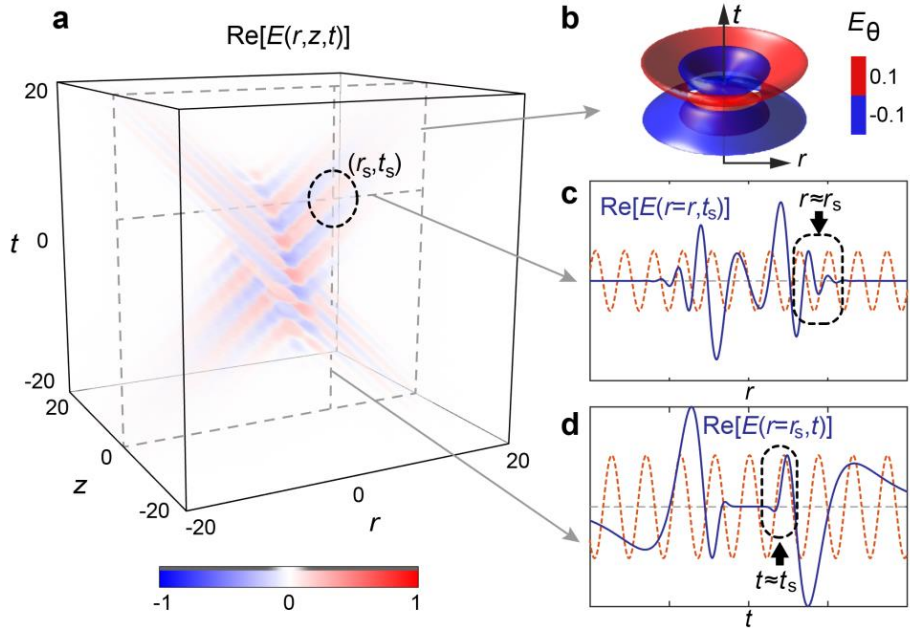

Figure 1: Space and time superoscillations in a band-limited supertoroidal light pulse ($\alpha=50$, $q_2=50q_1$). **a,** Spatiotemporal evolution of the azimuthal electric field, $E_\theta(r,z,t)$. **b,** Isosurfaces of the electric field, $E_\theta(r,z=0,t)$, in the focal plane. **c,** Radial profile of the electric field at focus ($z=0$) and at specific moment in time $t_s=5$ (blue line), and the corresponding fastest spatial frequency component (red-dashed line). The black dashed rectangle highlights the region spatial superoscillation ($r\sim r_s$). **d,** Temporal profile of the electric field at focus ($z=0$) and at specific radial position $r_s=10$ (blue line), and the corresponding fastest temporal frequency component (red-dashed line). The black dashed rectangle highlights the region spatial superoscillation ($t\sim t_s$).

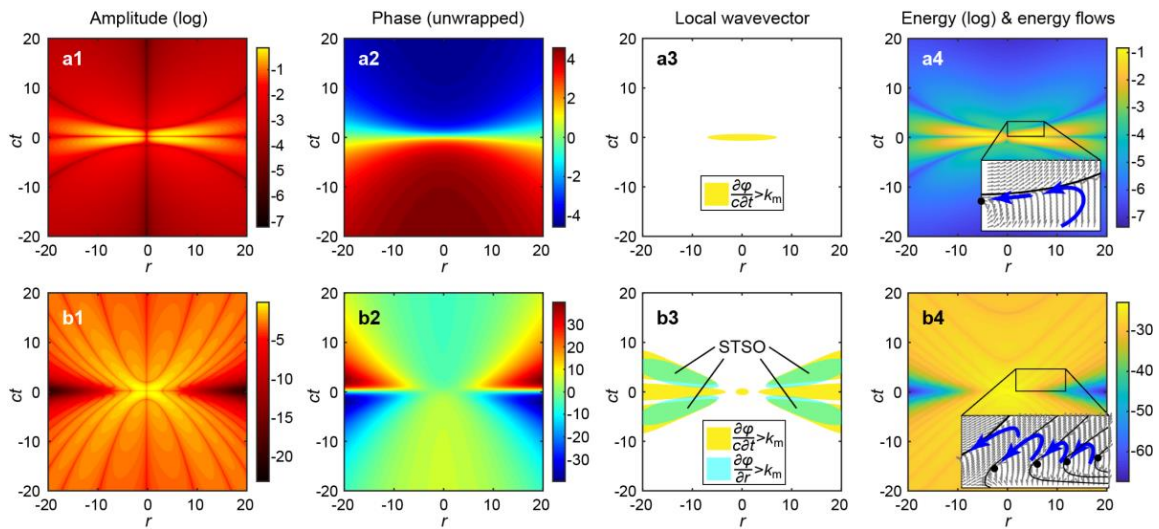

Figure 2: **a1,b1,** Spatiotemporal field modulus ($|E_\theta(r,z=0,t)|$) distributions for a toroidal pulse ($\alpha=1$) (**a1**) and an STP ($\alpha=50$) (**b1**), at focus ($z=0$). The amplitude is presented in terms of the logarithm of its real part, $\log_{10}|\mathrm{Re}[E_\theta(r,z=0,t)]|$. **a2,b2**, Unwrapped phase distributions $\varphi(r,t)=\mathrm{Arg}[E_\theta(r,z=0,t)]$ of the two pulses presented in (a1,b1). **a3,b3**, Regions in which the radial local wavevector ($\partial\varphi/\partial r$) and local temporal frequency ($\partial\varphi/\partial t$) of the toroidal (a3) and supertoroidal (b3) pulse exceed the threshold frequency, $\omega_m$, and wavevector, $k_m$, respectively ($k_m=2/q_1$). **a4,b4**, The energy spatial and temporal distribution, $w=(\epsilon_0 E^2+\mu_0 H^2)/2$, for the two pulses. Insets show the local energy flow with the black solid lines and dots marking the zero lines and singularities, and blue thick arrows marking the areas of energy backflow. Unit for all axes, $r$ and $ct$, is $q_1$.

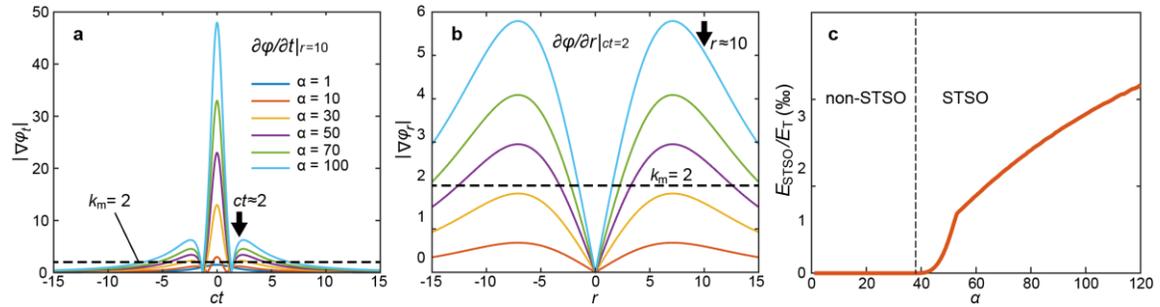

Figure 3: **a,b**, The temporal local frequency at $r=10$ (**a**) and the radial local wavevector at $t/c=2$ (**b**) of supertoroidal pulses for different values of $\alpha$. The black dashed lines mark the value of the fastest spatial and temporal frequency component, respectively. **c**, the ratio of energy in the STSO region over the total energy of the pulse as a function of supertoroidal order $\alpha$.

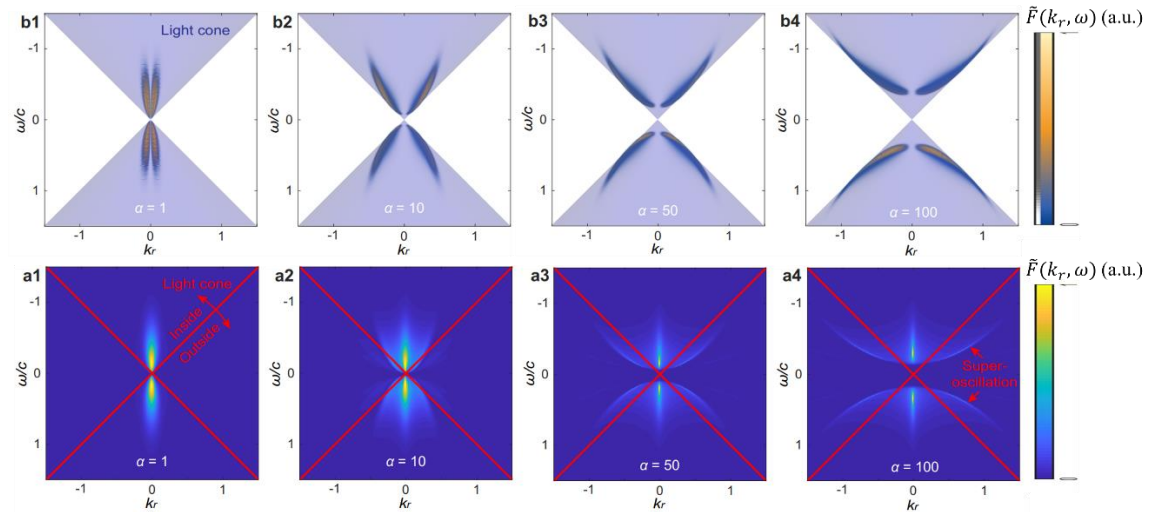

Figure 4: **a1-a4**, Spectral power projected in the $(k_r,\omega)$ plane for toroidal pulses ($\alpha=1$) and STPs ($\alpha=10, 50, 100$). The light blue regions mark the light cone. **b1-b4**, Spectral power projected in the $(k_r,\omega)$ plane corresponding to local STSO segments of pulses of different order ($\alpha=1, 10, 50, 100$). The red lines mark the boundary of the light cone. Note that, whereas the spectra of the full pulses (a1-a4) are confined on the surface of the light cone, local segments (b1-b4) exhibit out-of-cone components.

Supplementary Materials for

# Simultaneous Superoscillations in Space and Time in Nonseparable Light Pulses

Yijie Shen[1,2], Nikitas Papasimakis[3], Nikolay I. Zheludev[3,4]

1. Centre for Disruptive Photonic Technologies, School of Physical and Mathematical Sciences, Nanyang Technological University, Singapore 637371, Singapore
2. School of Electrical and Electronic Engineering, Nanyang Technological University, Singapore 639798, Singapore
3. Optoelectronics Research Centre & Centre for Photonic Metamaterials, University of Southampton, Southampton SO17 1BJ, UK
4. Institute for Advanced Study, Texas A&M University, USA

**Supplementary text**

**S1. Space-time solution of supertoroidal pulses**

To obtain the exact space-time nonseparable solution of electromagnetic waves, we start from the scalar generating function derived from modified power spectrum method [1,2]:

$$f = f_0 \frac{e^{-s/q_3}}{(q_1 + i\tau)(s + q_2)^\alpha} \tag{S1}$$

Where $s \equiv r^2/(q_1 + i\tau) - i\sigma$, $\tau = z - ct$, $\sigma = z + ct$, $r = \sqrt{x^2 + y^2}$, $(x, y, z)$ is the spatial coordinate, $t$ is time, $c = 1/\sqrt{\mu_0 \varepsilon_0}$ is the speed of light, the parameters $q_1, q_2, q_3$ are real positive with units of length, and the real dimensionless parameter $\alpha$ must satisfy $\alpha \geq 1$ in order for the electromagnetic pulse to fulfill finite energy. In conventional method, it always assumes that $q_3 \to \infty$ and $\alpha = 1$. Here we break the limit of the parameter $\alpha$, that can be any real number no less than one in our new derivation. The scalar generating function is given by:

$$\begin{aligned} f &= f_0 \frac{1}{(q_1 + i\tau)\left(\dfrac{r^2}{q_1 + i\tau} - i\sigma + q_2\right)^\alpha} \\ &= f_0 \frac{(q_1 + i\tau)^{\alpha-1}}{\left[r^2 + (q_1 + i\tau)(q_2 - i\sigma)\right]^\alpha} \end{aligned} \tag{S2}$$

The doughnut-like pulse is derived under a curled vector Hertz potential $\mathbf{\Pi} = \nabla \times \hat{\mathbf{z}} f(\mathbf{r}, t)$ in cylindrical coordinate $(r, \theta, z)$, the TE-mode electromagnetic field can be generated from Hertz potential by

$$\begin{cases} \mathbf{E}(\mathbf{r}, t) = -\mu_0 \dfrac{\partial}{\partial t} \nabla \times \mathbf{\Pi} = \hat{\boldsymbol{\theta}} \mu_0 \partial_r \partial_t f \\ \mathbf{H}(\mathbf{r}, t) = \nabla \times (\nabla \times \mathbf{\Pi}) = \hat{\mathbf{r}} \partial_r \partial_z f + \hat{\mathbf{z}} \left(\partial_z^2 - \dfrac{1}{c^2} \partial_t^2\right) f \end{cases} \tag{S3}$$

For this pulse solution, the electric field is purely azimuthally polarized, and the magnetic field is along the radial and longitudinal directions with no azimuthal component. The azimuthally polarized electric field of supertoroidal pulses is derived as [3]:

$$E = \mu_0 \partial_z \partial_t f = \mu_0 f_0 \partial_\rho \partial_t \frac{(q_1 + i\tau)^{\alpha-1}}{\left[r^2 + (q_1 + i\tau)(q_2 - i\sigma)\right]^\alpha}$$

$$= \mu_0 f_0 \left\{ \frac{-2\alpha(\alpha+1)icr(q_1+i\tau)^{\alpha-1}(q_1+q_2-2ict)}{\left[r^2+(q_1+i\tau)(q_2-i\sigma)\right]^{\alpha+2}} + \frac{2(\alpha-1)\alpha icr(q_1+i\tau)^{\alpha-2}}{\left[r^2+(q_1+i\tau)(q_2-i\sigma)\right]^{\alpha+1}} \right\} \quad (S4)$$

$$= f_0 \sqrt{\frac{\mu_0}{\varepsilon_0}} \left\{ \frac{-2\alpha(\alpha+1)ir(q_1+i\tau)^{\alpha-1}(q_1+q_2-2ict)}{\left[r^2+(q_1+i\tau)(q_2-i\sigma)\right]^{\alpha+2}} + \frac{2(\alpha-1)\alpha ir(q_1+i\tau)^{\alpha-2}}{\left[r^2+(q_1+i\tau)(q_2-i\sigma)\right]^{\alpha+1}} \right\}$$

This equation (S4) is the expression of high-order flying doughnut pulse. Note that when $\alpha=1$, the equation (S4) will be reduced into the expression of fundamental flying doughnut pulse [2]. For the paraxial limit condition, $q_2 \gg q_1$, and neglecting extremely small value of the second term, the equation (S4) can be simplified as:

$$E = -2\alpha(\alpha+1)if_0 \sqrt{\frac{\mu_0}{\varepsilon_0}} \frac{r(q_1+i\tau)^{\alpha-1}(q_2-2ict)}{\left[r^2+(q_1+i\tau)(q_2-i\sigma)\right]^{\alpha+2}} \quad (S5)$$

Considering the field is a very short propagating pulse at the speed of $c$, and $\tau = z - ct$ represents the local time, we can use the approximation of $z \doteq ct$ to evaluate the $\sigma = z + ct \doteq 2z$, then the electric field can be derived as:

$$E = -2\alpha(\alpha+1)if_0 \sqrt{\frac{\mu_0}{\varepsilon_0}} \frac{r(q_1+i\tau)^{\alpha-1}(q_2-i2z)}{\left[r^2+(q_1+i\tau)(q_2-i2z)\right]^{\alpha+2}}$$

$$= -2\alpha(\alpha+1)if_0 \sqrt{\frac{\mu_0}{\varepsilon_0}} r(q_1+i\tau)^{\alpha-1} \frac{1}{(q_2-i2z)^{\alpha+1}} \frac{1}{\left[\frac{r^2}{q_2-i2z} + (q_1+i\tau)\right]^{\alpha+2}}$$

$$= -2\alpha(\alpha+1)if_0 \sqrt{\frac{\mu_0}{\varepsilon_0}} r(q_1+i\tau)^{\alpha-1} \left(\frac{q_2+i2z}{4z^2+q_2^2}\right)^{\alpha+1} \frac{1}{\left[q_1 + \frac{r^2(q_2+i2z)}{4z^2+q_2^2} + i\tau\right]^{\alpha+2}} \quad (S6)$$

$$= -2\alpha(\alpha+1)if_0 \sqrt{\frac{\mu_0}{\varepsilon_0}} r(q_1+i\tau)^{\alpha-1} \left(\frac{q_2+i2z}{4z^2+q_2^2}\right)^{\alpha+1} \frac{1}{\left[q_1 + \frac{q_2 r^2}{4z^2+q_2^2} + i\left(\tau + \frac{2zr^2}{4z^2+q_2^2}\right)\right]^{\alpha+2}}$$

Here we define the notations of radius of curvature, $R(z)$, and beam waist profile, $w(z)$, as

$$R(z) = z\left[1 + \left(\frac{z_0}{z}\right)^2\right] = \frac{4z^2 + q_2^2}{4z} \quad (S7)$$

$$w^2(z) = w_0^2 \left[1 + \left(\frac{z}{z_0}\right)^2\right] = \frac{q_1}{2q_2}\left(4z^2 + q_2^2\right) \quad (S8)$$

with the Rayleigh length and basic waist constant given by $z_0 = \dfrac{q_2}{2}$ and $w_0^2 = \dfrac{q_1 q_2}{2} = q_1 z_0$.

Substitute these two notations and equations (S7) and (S8) into equation (S6), we can simplify electric field expression as:

$$
\begin{aligned}
E &= -2\alpha(\alpha+1) i f_0 \sqrt{\dfrac{\mu_0}{\varepsilon_0}} r(q_1+i\tau)^{\alpha-1} \dfrac{\left[q_2\left(1+i\dfrac{2z}{q_2}\right)\right]^{\alpha+1}}{\dfrac{2q_2}{q_1} w^2} \dfrac{1}{\left[q_1\left(1+\dfrac{r^2}{2w^2}\right)+i\left(\tau+\dfrac{r^2}{2R}\right)\right]^{\alpha+2}} \\
&= -2\alpha(\alpha+1) i f_0 \sqrt{\dfrac{\mu_0}{\varepsilon_0}} r(q_1+i\tau)^{\alpha-1} \dfrac{\left[w_0^2\left(1+i\dfrac{z}{z_0}\right)\right]^{\alpha+1}}{2z_0 w^2} \dfrac{1}{\left[q_1\left(1+\dfrac{r^2}{2w^2}\right)+i\left(\tau+\dfrac{r^2}{2R}\right)\right]^{\alpha+2}} \quad (S9)\\
&= -\dfrac{\alpha(\alpha+1)}{2^\alpha} i f_0 \sqrt{\dfrac{\mu_0}{\varepsilon_0}} r(q_1+i\tau)^{\alpha-1} \dfrac{w_0^{2(\alpha+1)}}{z_0^{\alpha+1} w^{2(\alpha+1)}} \left(1+i\dfrac{z}{z_0}\right)^{\alpha+1} \dfrac{1}{\left[q_1\left(1+\dfrac{r^2}{2w^2}\right)+i\left(\tau+\dfrac{r^2}{2R}\right)\right]^{\alpha+2}}
\end{aligned}
$$

Applying Taylor expansion of $\sqrt{1+x^2}\exp(i\tan^{-1}x) = 1+ix+O\left(\dfrac{x^2}{\sqrt{2}}\right)$, i.e. the approximation of $1+ix \doteq \sqrt{1+x^2}\exp(i\tan^{-1}x)$, the numerator term in equation (S9) can be rewritten as:

$$
\left(1+i\dfrac{z}{z_0}\right)^{\alpha+1} = \left\{\sqrt{1+\left(\dfrac{z}{z_0}\right)^2}\exp\left[i\cdot\tan^{-1}\left(\dfrac{z}{z_0}\right)\right]\right\}^{\alpha+1} = \dfrac{w^{\alpha+1}}{w_0^{\alpha+1}}\exp\left[i(\alpha+1)\phi(z)\right] \quad (S10)
$$

Where the $\phi(z) = \tan^{-1}\left(\dfrac{z}{z_0}\right)$ is the Gouy phase. Substitute equation (S10) into (S9) to carry on the simplification:

$$
\begin{aligned}
E &= -\dfrac{\alpha(\alpha+1)}{2^\alpha} i f_0 \sqrt{\dfrac{\mu_0}{\varepsilon_0}} r(q_1+i\tau)^{\alpha-1} \dfrac{w_0^{2(\alpha+1)}}{z_0^{\alpha+1} w^{2(\alpha+1)}} \exp\left[i(\alpha+1)\phi(z)\right] \cdot \dfrac{1}{\left[q_1\left(1+\dfrac{r^2}{2w^2}\right)+i\left(\tau+\dfrac{r^2}{2R}\right)\right]^{\alpha+2}} \\
&= -\dfrac{\alpha(\alpha+1)}{2^\alpha} i f_0 \sqrt{\dfrac{\mu_0}{\varepsilon_0}} \dfrac{w_0^{\alpha+1} r(q_1+i\tau)^{\alpha-1}}{z_0^{\alpha+1} w^{\alpha+1}} \cdot \dfrac{\left[q_1\left(1+\dfrac{r^2}{2w^2}\right)-i\left(\tau+\dfrac{r^2}{2R}\right)\right]^{\alpha+2}}{\left[\left(q_1\left(1+\dfrac{r^2}{2w^2}\right)\right)^2+\left(\tau+\dfrac{r^2}{2R}\right)^2\right]^{\alpha+2}} \cdot \exp\left[i(\alpha+1)\phi(z)\right]
\end{aligned}
$$

(S11)

Define the notations of radially scaled local time, $T(\mathbf{r},\tau)$, as

$$T = \frac{-\left(\tau + \dfrac{r^2}{2R}\right)}{q_1\left(1+\dfrac{r^2}{2w^2}\right)} = \frac{c\left(t - \dfrac{z+r^2/2R}{c}\right)}{q_1\left(1+\dfrac{r^2}{2w^2}\right)} \tag{S12}$$

Substitute equation (S12) into (S11) to carry on the simplification:

$$\begin{aligned}
E &= -\alpha(\alpha+1)if_0\sqrt{\frac{\mu_0}{\varepsilon_0}}\frac{r(q_1+i\tau)^{\alpha-1}}{2^\alpha\left(z^2+z_0^2\right)^{(\alpha+1)/2}} \cdot \frac{\left[q_1\left(1+\dfrac{r^2}{2w^2}\right)\right]^{\alpha+2}(1+iT)^{\alpha+2}}{\left[\left(q_1\left(1+\dfrac{r^2}{2w^2}\right)\right)^2(1+T^2)\right]^{\alpha+2}} \cdot \exp\left[i(\alpha+1)\phi(z)\right] \\
&= -\frac{\alpha(\alpha+1)}{2^\alpha}if_0\sqrt{\frac{\mu_0}{\varepsilon_0}}\frac{w_0^{\alpha+1}r(q_1+i\tau)^{\alpha-1}}{z_0^{\alpha+1}w^{\alpha+1}} \cdot \frac{(1+iT)^{\alpha+2}}{\left(q_1\left(1+\dfrac{r^2}{2w^2}\right)\right)^{\alpha+2}(1+T^2)^{\alpha+2}} \cdot \exp\left[i(\alpha+1)\phi(z)\right]
\end{aligned}$$

$$\tag{S13}$$

Applying the Taylor approximation of equation to simplify the numerator term in equation (S13), we get:

$$(1+iT)^{\alpha+2} \doteq \left[\sqrt{1+T^2}\exp\left(i\tan^{-1}T\right)\right]^{\alpha+2} = (1+T^2)^{(\alpha+2)/2}\exp\left[i\cdot(\alpha+2)\tan^{-1}T\right] \tag{S14}$$

To simplify the numerator term in equation (S13), we carry on the simplification of the electric field expression:

$$\begin{aligned}
E &= -\frac{\alpha(\alpha+1)}{2^\alpha}if_0\sqrt{\frac{\mu_0}{\varepsilon_0}}\frac{w_0^{\alpha+1}r(q_1+i\tau)^{\alpha-1}}{z_0^{\alpha+1}w^{\alpha+1}} \cdot \frac{(1+T^2)^{(\alpha+2)/2}\exp\left[i\cdot(\alpha+2)\tan^{-1}T\right]}{\left(q_1\left(1+\dfrac{r^2}{2w^2}\right)\right)^{\alpha+2}(1+T^2)^{\alpha+2}} \cdot \exp\left[i(\alpha+1)\phi(z)\right] \\
&= -\frac{\alpha(\alpha+1)}{2^\alpha}if_0\sqrt{\frac{\mu_0}{\varepsilon_0}}\frac{w_0^{\alpha+1}r(q_1+i\tau)^{\alpha-1}}{z_0^{\alpha+1}w^{\alpha+1}\left(q_1\left(1+\dfrac{r^2}{2w^2}\right)\right)^{\alpha+2}(1+T^2)^{(\alpha+2)/2}} \cdot \exp\left\{i\left[(\alpha+2)\tan^{-1}T+(\alpha+1)\phi(z)\right]\right\}
\end{aligned}$$

$$\tag{S15}$$

The numerator term in equation (S15) can be further simplified by using the Taylor approximation to separate the amplitude and phase terms as:

$$\begin{aligned}
(q_1+i\tau)^{\alpha-1} &= \left[q_1\left(1+i\frac{\tau}{q_1}\right)\right]^{\alpha-1} = \left(q_1\sqrt{1+\left(\frac{\tau}{q_1}\right)^2}\exp\left[i\tan^{-1}\left(\frac{\tau}{q_1}\right)\right]\right)^{\alpha-1} \\
&= (q_1^2+\tau^2)^{(\alpha-1)/2}\exp\left[i(\alpha-1)\tan^{-1}\left(\frac{\tau}{q_1}\right)\right]
\end{aligned} \tag{S16}$$

Substitute equation (S16) into (S15) to carry on the derivation:

$$E = -\frac{\alpha(\alpha+1)}{2^{\alpha}} i f_0 \sqrt{\frac{\mu_0}{\varepsilon_0}} \frac{w_0^{\alpha+1} r \left(q_1^2 + \tau^2\right)^{(\alpha-1)/2}}{z_0^{\alpha+1} w^{\alpha+1} \left[q_1\left(1+\frac{r^2}{2w^2}\right)\right]^{\alpha+2} \left(1+T^2\right)^{(\alpha+2)/2}} \cdot$$
$$\times \exp\left\{i\left[(\alpha-1)\tan^{-1}\left(\frac{\tau}{q_1}\right) + (\alpha+2)\tan^{-1}T + (\alpha+1)\phi(z)\right]\right\}$$
(S17)

Define the notations of generalized local-time amplitude, $A_\alpha(\mathbf{r},\tau)$, and generalized local-time wavenumber, $k_\alpha(\mathbf{r},\tau)$, as

$$A_\alpha(\mathbf{r},\tau) = -f_0 \sqrt{\frac{\mu_0}{\varepsilon_0}} \frac{\left(q_1^2+\tau^2\right)^{(\alpha-1)/2}}{q_1^{\alpha+2}(T^2+1)^{(\alpha+2)/2}} = \frac{-f_0\mu_0 c \left(q_1^2+\tau^2\right)^{(\alpha-1)/2}}{q_1^{\alpha+2}(T^2+1)^{(\alpha+2)/2}}$$
(S18)

$$k_\alpha(\mathbf{r},\tau) = (\alpha-1)\tan^{-1}\left(\frac{\tau}{q_1}\right) + (\alpha+2)\tan^{-1}T$$
(S19)

Substitute equations (S18) and (S19) into (S17), the final closed-form amplitude-phase expression of supertoroidal pulses is given by:

$$E = i \frac{\alpha(\alpha+1) w_0^{\alpha+1} r A_\alpha(\mathbf{r},\tau)}{2^\alpha z_0^{\alpha+1} w^{\alpha+1} \left(1+\frac{r^2}{2w^2}\right)^{\alpha+2}} \exp\left\{i[k_\alpha(\mathbf{r},\tau) + (\alpha+1)\phi(z)]\right\}$$
(S20)

The distribution of amplitude $|E|$ and phase $\mathrm{Arg}(E)$ of supertoroidal pulse can be calculated from this equation. With the similar derivation, the expressions for the magnetic field, including radial and longitudinal components, can be given as:

$$H_r = \partial_r \partial_z f = f_0 \partial_r \partial_z \frac{(q_1+i\tau)^{\alpha-1}}{\left[r^2 + (q_1+i\tau)(q_2-i\sigma)\right]^\alpha}$$
$$= f_0\left\{\frac{2\alpha(\alpha+1)ir(q_1+i\tau)^{\alpha-1}(q_2-q_1-2iz)}{\left[r^2+(q_1+i\tau)(q_2-i\sigma)\right]^{\alpha+2}} - \frac{2(\alpha-1)\alpha ir(q_1+i\tau)^{\alpha-2}}{\left[r^2+(q_1+i\tau)(q_2-i\sigma)\right]^{\alpha+1}}\right\}$$
(S21)

$$H_z = \left(\partial_z^2 - \frac{1}{c^2}\partial_t^2\right)f = f_0\left(\partial_z^2 - \frac{1}{c^2}\partial_t^2\right)\frac{(q_1+i\tau)^{\alpha-1}}{\left[r^2+(q_1+i\tau)(q_2-i\sigma)\right]^\alpha}$$
$$= f_0\left\{\frac{-4\alpha(q_1+i\tau)^{\alpha-1}\left[r^2-\alpha(q_1+i\tau)(q_2-i\sigma)\right]}{\left[r^2+(q_1+i\tau)(q_2-i\sigma)\right]^{\alpha+2}} - \frac{4(\alpha-1)\alpha(q_1+i\tau)^{\alpha-2}(q_2-i\sigma)}{\left[r^2+(q_1+i\tau)(q_2-i\sigma)\right]^{\alpha+1}}\right\}$$
(S22)

and the radial magnetic component can be derived into the amplitude-phase formation

$$H_r = -i\sqrt{\frac{\varepsilon_0}{\mu_0}} \frac{\alpha(\alpha+1)w_0^{\alpha+1}rA_\alpha(\mathbf{r},\tau)}{2^\alpha z_0^{\alpha+1}w^{\alpha+1}\left(1+\frac{r^2}{2w^2}\right)^{\alpha+2}} \exp\{i[k_\alpha(\mathbf{r},\tau)+(\alpha+1)\phi(z)]\} \quad (S23)$$

Based on the transverse components of the electromagnetic field, Eqs. (S20) and (S23), The transverse electromagnetic field can reach a more compact formation if we create a notation of the complex amplitude of $A_\alpha = i\dfrac{\alpha(\alpha+1)w_0^{\alpha+1}rA_\alpha(\mathbf{r},\tau)}{2^\alpha z_0^{\alpha+1}w^{\alpha+1}\left(1+r^2/(2w^2)\right)^{\alpha+2}}$, the unified expression of the transverse field can be simplified as:

$$\Psi_\perp = \begin{bmatrix} \mathbf{E}_\perp \\ \mathbf{H}_\perp \end{bmatrix} = \begin{bmatrix} \hat{\boldsymbol{\theta}} \\ \sqrt{\frac{\varepsilon_0}{\mu_0}}\hat{\mathbf{r}} \end{bmatrix} A_\alpha \exp\{i[k_\alpha(\mathbf{r},\tau)+(\alpha+1)\phi(z)]\} \quad (C12)$$

## S2. Space-Time Superoscillations in band-limited STPs

The spectrum of a supertoroidal light pulse (STP), given by $E(r, z, t)$, can be directly obtained by Fourier transformation from space-time $(r, z, t)$ to spatiotemporal frequency $(f_r, f_z, f)$ domain.

$$\tilde{E}(f_r, f_z, f) = \int_{-\infty}^{\infty}\int_{-\infty}^{\infty}\int_{-\infty}^{\infty} E(r,z,t)\exp\left[-i2\pi(f_r r + f_z z + ft)\right]drdzdt \quad (S21)$$

Limited by the solution of Maxwell's equations, the entire spectrum of a space-time light pulse must be distributed on the surface of light cone, i.e the conic surface with unit slope of its generatrix in the coordinate $(k_r, k_z, \omega/c)$, where the spatial wavevectors are related to the spatial frequencies by $k_r=2\pi f_r/c$ and $k_r=2\pi f_z/c$, and temporal frequency is related to the angular frequency by $\omega=2\pi f$. In numerical calculation, we should make sure the range of space-time is set large enough to cover effective energy distribution (the energy at the boundary is 1/1000 less than the entire energy). Figure S1 shows the theoretical results of the spectra of several supertoroidal pulses with various of orders. Although the entire spectrum of a supertoroidal pulse must be located on the light cone, but the local spectrum of which does not necessary to fulfill this requirement and can be distributed over the limit of light cone, for example, the radial-temporal spectrum of a supertoroidal pulse at a specific propagation distance:

$$\tilde{E}(f_r, f) = \int_{-\infty}^{\infty}\int_{-\infty}^{\infty} E(r,0,t)\exp\left[-i2\pi(f_r r + ft)\right]drdt \quad (S22)$$

as the results of Figure 3 in the main text show.

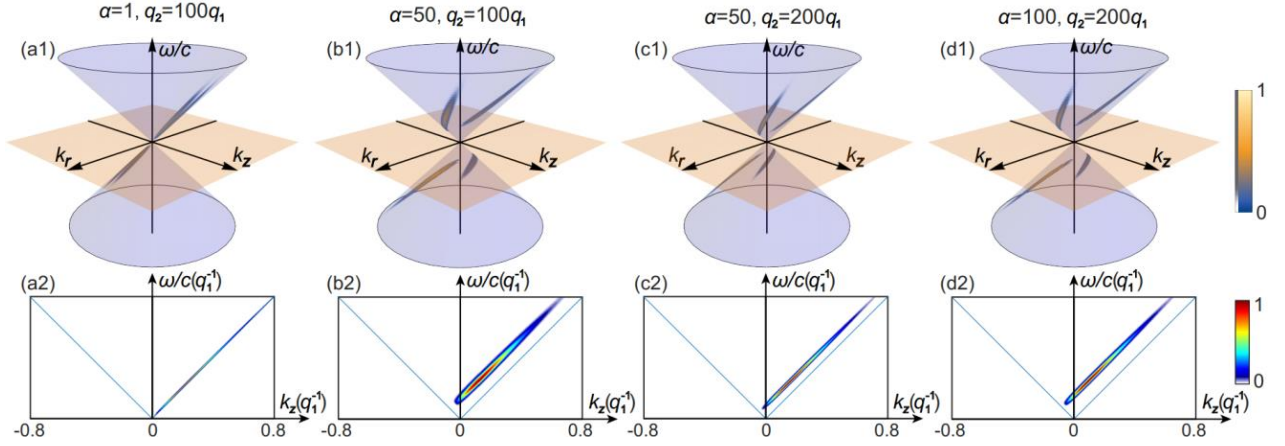

Figure S1: (a1-d1) The power spectra of various supertoroidal pulses in ($k_r$, $k_z$, $\omega/c$) domain, and (a2-d2) their projection patterns onto the $k_z-\omega/c$ plane. The blue surface is the light cone. Parameters: (a) $\alpha=1$, $q_2=100q_1$; (b) $\alpha=50$, $q_2=100q_1$; (c) $\alpha=50$, $q_2=200q_1$; (d) $\alpha=100$, $q_2=200q_1$.

Theoretically, an ideal supertoroidal pulse is not exactly band-limited (boundary by zero-value), while the effective energy is localized in a finite region (exponentially decayed toward infinity). Therefore, we can still find the effective band-limited effect in such pulses. In the numerical calculation of results of Figure S1, the band limits for the $k_r$, $k_z$, $\omega/c$ are all set as 2.5, which will not impact on the reconstruction of spatiotemporal signal (inverse Fourier Transform):

$$\tilde{\tilde{E}}(r,z,t) = \int_{-\infty}^{\infty}\int_{-\infty}^{\infty}\int_{-\infty}^{\infty} \tilde{E}(f_r, f_z, f)\exp\left[i2\pi(f_r r + f_z z + ft)\right] df_r df_z df \tag{S23}$$

We compared the normalized reconstructed signal and theoretical signal and make sure the maximal amplitude difference between them is less than 0.01. In this case, the band-limited effect cannot be revealed in numerical calculation. In order to reveal the band-limited effect and study superoscillation effect, we proposed an approach of on-demand band-limited modulation onto a supertoroidal pulse, which is illustrated in Figure S2. We start from an ideal space-time signal of supertoroidal pulse (Figure S2(a)) and the entire spectrum can be numerical simulated by Fourier transformation (Figure S2(b)). In order to obtain a band-limited signal, we use the product of the entire spectrum and a frequency dependent bump function to be as the spectrum of the modulated signal. The bump function is a continue function between 0 and 1, which can be defined as

$$B(f) = \begin{cases} 1, & |f| < f_m - \delta \\ \exp\left[-\dfrac{1}{1-(f \mp f_m)^2/\delta^2}\right], & f_m - \delta < \pm f < f_m \\ 0, & \text{otherwise} \end{cases} \tag{S24}$$

Such a bump function is plotted in Figure S2(c), where $f_m$ is the cutting-off frequency to set exact band limit and $\delta$ is a small frequency gap to ensure a physical smooth change of function (avoid non-physical step). Then the band-limited spectrum of the modulated signal can be expressed as:

$$\tilde{E}_c(f_r, f_z, f) = \tilde{E}(f_r, f_z, f)B(f) \tag{S25}$$

Figure S2(d) shows a simulated result of the band-limited spectrum with cutting-off frequency of $f_m = 0.6\ (c/q_1)$ and plotted onto the light cone. In our simulation, the $\delta$ is set as $f_m/20$. The corresponding band-limited spatiotemporal supertoroidal-like pulse can be obtained by inverse Fourier transform:

$$\tilde{\tilde{E}}_c(r,z,t) = \int_{-\infty}^{\infty}\int_{-\infty}^{\infty}\int_{-\infty}^{\infty} \tilde{E}_c(f_r, f_z, f)\exp\left[i2\pi(f_r r + f_z z + ft)\right]df_r df_z df \tag{S26}$$

Such a synthetic spatiotemporal pulse, we call band-limited supertoroidal-like pulse (arbitrary value of supertoroidal order $\alpha$ can be set in which), has exact band limit of spatial and temporal frequencies $(f_r, f_z, f)$ with value of $f_m$, that can be arbitrarily tuned.

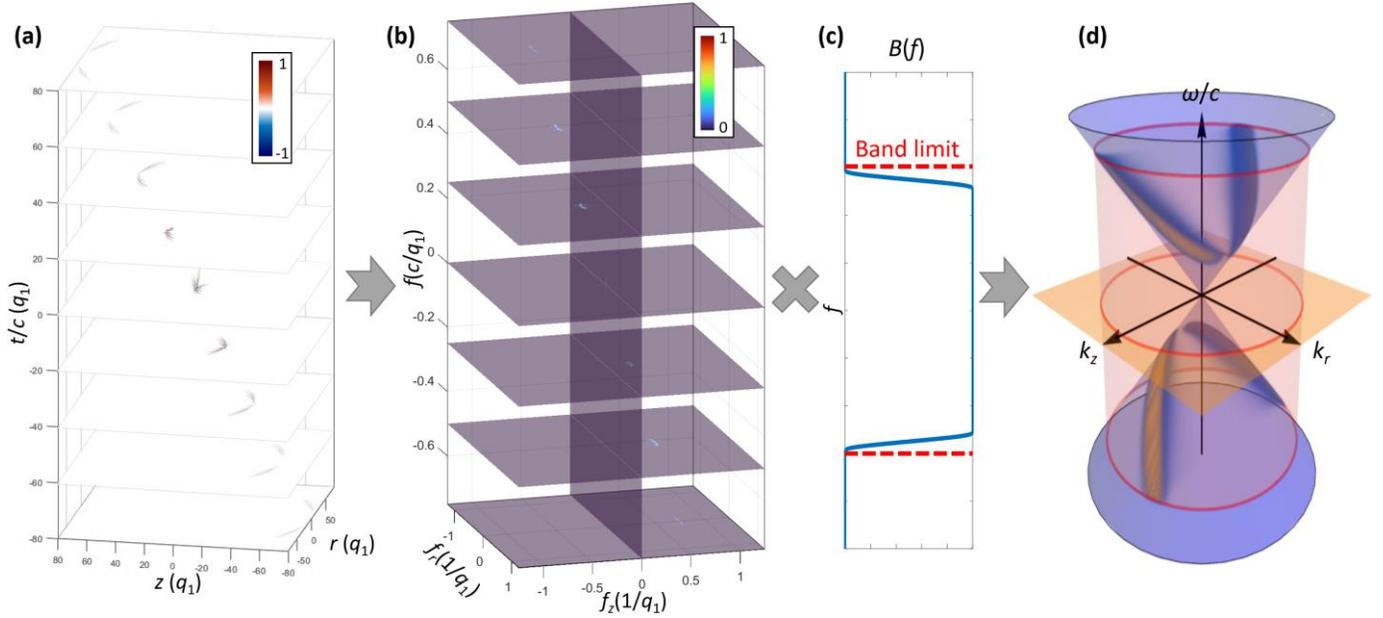

Figure S2: The process of setting band-limited modulation onto a supertoroidal pulse. (a) The space-time distributions of the amplitude of a supertoroidal pulse ($\alpha=20$, $q_2=50q_1$), and (b) its numerically calculated entire spectrum. (c) The frequency dependent bump function with on-demand band-limit set. (d) The band-limited entire spectrum is obtained by multiply the original entire spectrum and the bump function, which is plotted on a light cone with red line highlighting the band limit.

In order to qualitatively characterize the superoscillation of band-limited supertoroidal-like pulse, we numerically compared the oscillatory structures of the fastest locally oscillatory signal and the fastest harmonic Fourier component. In this process, we tune the parameters of cutting-off frequency and supertoroidal order to find the condition for space-time superoscillation.

For studying the spatial superoscillation, we construct a set of the spatiotemporal signal of band-limited supertoroidal-like pulses by equation (S26), for various values of cutting-off frequency $f_m$, supertoroidal order $\alpha$, and extract the radial oscillating curve at $t=0$ and $z=0$. Then, we find all the zero points and calculate the distance between each pair of adjacent zero point, and the minimal value of which corresponds to the fastest radial half-cycle local oscillation, as shown in Figure

S3. The fastest single-cycle radial local oscillation can be traced by similar way, i.e. the minimal distance across three adjacent zero points.

Similarly, for studying the temporal superoscillation, we construct a set of the spatiotemporal signal of band-limited supertoroidal-like pulses by equation (S26), for various values of cutting-off frequency $f_m$, supertoroidal order $\alpha$, and extract the temporal oscillating curve at $r=10q_1$ and $z=0$. Then, we find all the zero points and calculate the distance between each pair of adjacent zero point, and the minimal value of which corresponds to the fastest half-cycle temporal local oscillation. The fastest single-cycle temporal local oscillation can be traced by similar way, i.e. the minimal distance across three adjacent zero points.

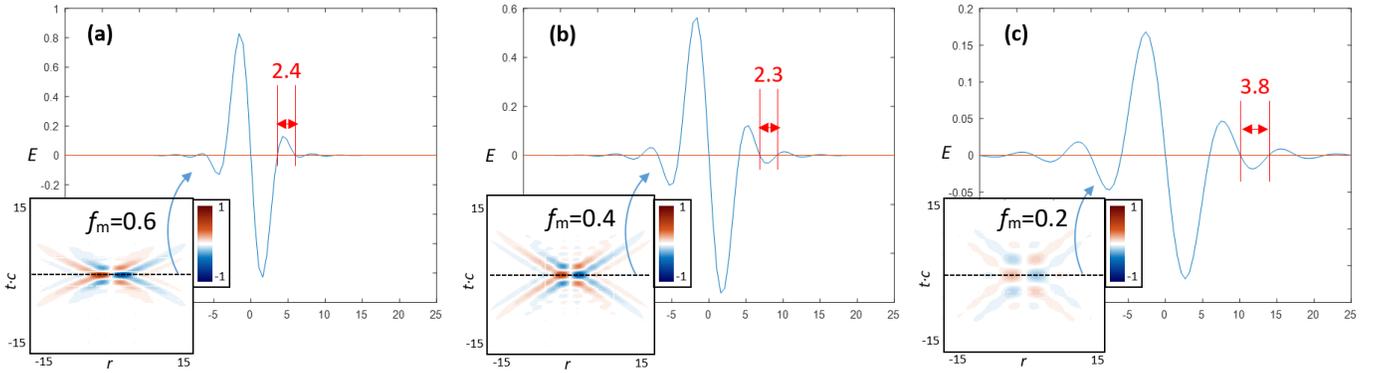

Figure S3: The examples of tracing the minimal radial oscillatory structures of band-limited supertoroidal-like pulses with cutting-off frequency set as (a) $f_m = 0.6$, (b) $f_m = 0.4$, (c) $f_m = 0.2$ (unit: $c/q_1$) at $t=0$ and $z=0$. The insets show corresponding space-time amplitude distributions. The results of minimal redial half-cycle oscillatory structure are marked in red with unit of $q_1$.

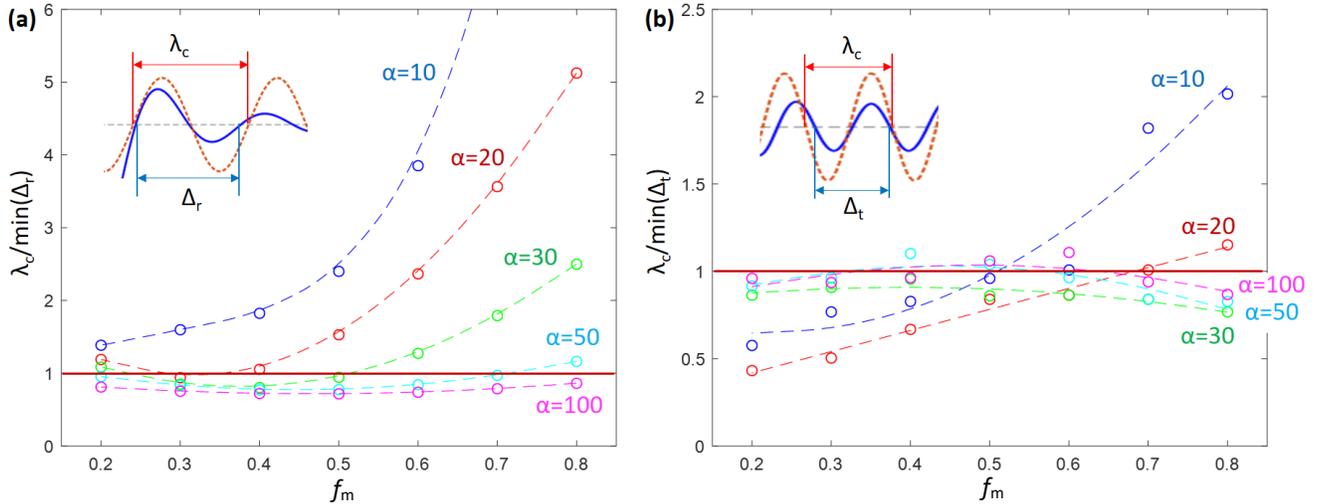

Figure S4: The results of (a) the ratio of cutting-off wavelength (of the fastest radial Fourier component) and the local minimal radial single-cycle oscillatory structure and (b) the ratio of cutting-off wavelength (of the fastest temporal Fourier component) and the local minimal temporal single-cycle oscillatory structure versus the cutting-off frequency $f_m$ ($f_m$=0.2, 0.3, 0.4, 0.5, 0.6, 0.7,

0.8, unit: $1/q_1$) for various values of supertoroidal order $α$ ($α=10$, $α=20$, $α=30$, $α=50$, $α=100$). The red lines mark the threshold of superoscillation.

In order to qualitatively characterize the spatial superoscillation, we calculate the cutting-off wavelength (of the fastest radial Fourier component), $λ_m=c/f_m$, and the local minimal radial single-cycle oscillatory structure, noted as $Δ_r$, versus the cutting-off frequency $f_m$ ($f_m=0.2$, 0.3, 0.4, 0.5, 0.6, 0.7, 0.8, unit: $1/q_1$) for various values of supertoroidal order $α$ ($α=10$, $α=20$, $α=30$, $α=50$, $α=100$), the results of which are shown in Figure S4(a). We can observe the evidence of spatial superoscillation that there are cases where $λ_m/\min(Δ_r)$ is less than 1. For a proper setting of cutting-off frequency, the higher value of supertoroidal order the easier to observe the spatial superoscillation (at the range we studied $α<100$). In order to qualitatively characterize the temporal superoscillation, we calculate the cutting-off wavelength (of the fastest temporal Fourier component), $λ_m=c/f_m$, and the local minimal temporal single-cycle oscillatory structure, noted as $Δ_t$, versus the cutting-off frequency $f_m$ ($f_m=0.2$, 0.3, 0.4, 0.5, 0.6, 0.7, 0.8, unit: $1/q_1$) for various values of supertoroidal order $α$ ($α=10$, $α=20$, $α=30$, $α=50$, $α=100$), the results of which are shown in Figure S4(b). We can observe the evidence of spatial superoscillation that there are cases where $λ_m/\min(Δ_t)$ is less than 1. For a proper setting of cutting-off frequency, the ratio is firstly decreased, over the threshold of superoscillation, then increased versus the increasing of supertoroidal order. In these results, we can find the cases where spatial superoscillation and temporal superoscillation occur simultaneously. For instance, the cases that $f_m=0.3$, 0.4, 0.5 and $α=20$, $f_m=0.2$, 0.6, 0.7, 0.8 and $α=50$, and $f_m=0.2$, 0.7, 0.8 and $α=100$. These are evidences of space-time superoscillation,